\documentclass[useAMS,usegraphicx,usenatbib]{mn2e}
\usepackage{times}

\begin{document}

\title[Hydrogenated protonated naphthalene and proflavine]{Electronic absorption spectra of hydrogenated protonated naphthalene and proflavine}
\author[A. Bonaca  and G. Bilalbegovi\' c]{A. Bonaca\thanks{Present address: Department of Astronomy, Yale University, New Haven, CT 06520-8101, USA}  and G. Bilalbegovi\' c\\
Department of Physics, Faculty of Science, University of Zagreb,
Bijeni{\v c}ka 32, 10000 Zagreb, Croatia}

\date{\today}

\pagerange{\pageref{firstpage}--\pageref{lastpage}}
\pubyear{2011} \volume{000}

\maketitle 
\label{firstpage}

\begin{abstract}
We study hydrogenated cations  of two polycyclic hydrocarbon molecules 
as models of hydrogenated organic species that form in the interstellar medium.
Optical spectra of  the hydrogenated naphthalene cation H$_n$-C$_{10}$H$_8$$^{+}$ for $n=1,2$, and $10$,
as well as of the astrobiologically interesting
hydrogenated proflavine cation H$_n$-C$_{13}$H$_{11}$N$_3$$^{+}$ for $n=1$ and $14$,
are calculated. The pseudopotential time dependent density functional theory is used.
It is found that the fully hydrogenated proflavine cation H$_{14}$-C$_{13}$H$_{11}$N$_3$$^{+}$ shows a broad spectrum where positions of individual
lines are almost lost.
The positions of lines, their shapes, and intensities change in hydronaphthalene and hydroproflavine cations showing that
hydrogen additions  induce substantially different optical spectra in a comparison to  base polycyclic hydrocarbon cations.
One calculated line in the visible spectrum of H$_{10}$-C$_{10}$H$_8$$^{+}$, and one in the
visible spectrum of H-C$_{13}$H$_{11}$N$_3$$^{+}$  
are close to the measured diffuse interstellar bands. We also present positions of near-UV lines.
\end{abstract}

\begin{keywords}
astrochemistry -- molecular processes -- methods:numerical
\end{keywords}

\section{Introduction}
\label{intro}

Hydrogen, in its atomic and molecular form, is the most abundant of chemical species in the interstellar medium. The number and size of detected interstellar molecules have been increasing last eighty years.
Polycyclic aromatic hydrocarbons (PAHs) have been investigated as carriers of the diffuse interstellar bands (DIBs), because of
unidentified infrared (UIR) emission features, and as a source of an anomalous microwave emission \citep{Tielens2008}.
It has been suggested that protonated PAH molecules play a catalytic role in the process of the H$_2$ formation in space \citep{Bauschlicher1998,Hirama2004}.
Electronic transitions in the optical spectral region of protonated hydrogenated coronene, ovalene, pyrene, and circumpyrene have been recently calculated and discussed in the context of the DIBs problem
\citep{Pathak2008,Hammonds2009}. Spectra of
isomers of protonated anthracene and phenanthrene have been measured in neon matrices and studied by the time-dependent density  functional theory
\citep{Garkusha2011}.
We have selected here cations  of naphthalene and proflavine  to study the change
of their overall optical spectra under additions of hydrogen atoms.
The presence of C$_{10}$H$_8$$^{+}$ in the interstellar medium has been discussed \citep{Iglesias-Groth2008,Galazutdinov2011}. 
Several times in the past the evidence of the interstellar naphthalene cation was reported and later shown to be premature 
\citep{Galazutdinov2011}.
Proflavine is a substituted PAH molecule  with possible applications in astrophysics and astrobiology \citep{Sarre2006,Bonaca2010}.

A  theoretical study of reactions  has shown that  the first H atom attaches to the naphthalene cation C$_{10}$H$_8^+$ in exothermic reaction of about 60 kcal mol$^{-1}$, whereas the process for the second H atom is exothermic by 45 kcal mol$^{-1}$ \citep{Bauschlicher1998}.
Herbst and Le Page have found that the naphthalene cation efficiently associates with atomic hydrogen at all densities \citep{Herbst1999}.
Calculations have also shown that the addition of H atoms to the naphthalene cation proceeds with little to no barriers
\citep{Ricca2007}. 
Electronic absorption spectra of
 H-C$_{10}$H$_8$$^{+}$ have been studied by the photofragment spectroscopy and theoretical analysis 
\citep{Alata2010a,Alata2010b}. It has been found that the protonated naphthalene cation absorbs in the visible part of the spectrum around 500 nm.
A possibility that C$_{10}$H$_8$ and C$_{10}$H$_8$$^{+}$ are carriers of 
DIBs has been studied \citep{Salama1992,Hirata1999,Krelowski2001,Malloci2007b}.
The evidence of the naphthalene cation  in the direction of the star Cernis 52 in the Perseus molecular cloud complex has been reported \citep{Iglesias-Groth2008}.
The presence of some known DIBs has been confirmed in this study, and two new bands consistent with laboratory measurements on the naphthalene cation have been observed. It has also been proposed that  hydrogen additions produce hydronaphthalene cations which contribute to
the anomalous microwave emission in this cloud \citep{Iglesias-Groth2008}. However, a recent work \citep{Galazutdinov2011} has shown that this report \citep{Iglesias-Groth2008} is premature.
The presence of the naphthalene cation and related species in the interstellar medium deserves further studies.

In addition to PAHs, the related molecules where C and H are substituted by other atoms are also studied in the astrophysical context \citep{Hudgins2005,Sarre2006}.
Nitrogen is abundant in the interstellar medium. Its compounds have been detected in interstellar dust particles and meteorites.
Organic dye proflavine, C$_{13}$H$_{11}$N$_3$ (3,6-diaminoacridine),
has been proposed as one of molecules which act as a molecular ``midwife'' because of their property to accelerate DNA and RNA synthesis \citep{Jain2004}. Therefore, the optical spectrum of proflavine and its ions is of astrobiological interest.
Studies of such large biological molecules are important for understanding of prebiotic chemistry \citep{Puletti2010}.
Organic dyes often exhibit strong lines in the visible spectrum.
It has been shown that the optical spectrum of proflavine in aqueous solutions strongly depends on the pH value \citep{DeSilvestri1984,Mennucci2005}. Protons can attach to nitrogen and carbon atoms in the molecule, and the state of proflavine in water and other solvents depends on pH of the solution.
For example, it has been measured that the maximum of absorption in the visible spectrum of proflavine ($\lambda_{max}$) in water at room temperature,
changes from 444 nm to 394 nm when pH changes from 7.0 to 14.0 \citep{DeSilvestri1984}.
 The visible and UV optical spectra of proflavine and its ions
have been recently studied using the pseudopotential time-dependent density functional theory methods \citep{Bonaca2010}. Positions of spectral lines have been compared with DIBs, but with no definite conclusions.
Because of the sensitivity of optical properties of proflavine in water on the pH factor, 
we investigated here the impact of hydrogen additions on the optical spectrum of its cation.

In this work we consider minimal and maximal hydrogenations of two organic cations.
An addition of one, two and ten H atoms to the naphthalene cation, as well as one (in two positions) and fourteen hydrogen atoms to the proflavine cation are studied.
We use the pseudopotential density functional theory (DFT) \citep{Martin2004} to determine the ground states of all base and hydrogenated species.
Calculations of optical spectra of these systems within the pseudopotential time-dependent density functional theory (TDDFT)
methods \citep{Runge1984} are described in section~\ref{methods}. Results are presented and
discussed in section~\ref{results}, while final remarks are given in section~\ref{concl}.

\section{Computational methods}
\label{methods}

Optical spectra are calculated using the Octopus code \citep{Castro2006}.
The geometries of all naphthalene and proflavine systems are minimized independently
by the Quantum ESPRESSO DFT package \citep{Giannozzi2009}. This approach is taken because of the fact that
the Octopus  is not in general suitable  for a search of optimal geometries.
Conditions as close as possible to simulations of the spectra within the Octopus code have been used in the Quantum ESPRESSO calculations
(i.e.,  the local density approximation (LDA)  and corresponding pseudopotentials, see below and in Bonaca \& Bilalbegovi\' c 2010).
The optimized geometries of cations are calculated without imposing any constraints, and they are
used as inputs in the ground state and time-dependent calculations in the Octopus code.

Optical spectra are obtained using a time propagation method and
the approximated enforced time-reversal symmetry algorithm \citep{Castro2004}. 
The ground state is perturbed by an impulsive time-dependent potential and 
the time evolution is followed for 15.8 fs. 
In the time propagation TDDFT method used in this work, the width of spectral lines  is inversely proportional to the total 
propagation time \citep{Castro2006,Lopez2003,Malloci2007b,Puletti2010}.
The step of 0.0012 $\hbar/$eV is applied.
The absolute cross-section  $\sigma(\omega)$ is obtained from dynamical polarisability $\alpha(\omega)$ which is calculated from the Fourier transform
of the time-dependent dipole momentum of the system.
Electronic spectra are calculated from
\begin{equation}
\sigma(\omega)=\frac{2\omega}{\pi}Im\; \alpha(\omega).
\label{cross}
\end{equation}
In this equation $Im\; \alpha(\omega)$ is the imaginary part of the dynamical polarisability.
The Troullier-Martins pseudopotentials \citep{Troullier1993} and the TDLDA approximation with the Perdew-Zunger exchange-correlation functional \citep{Perdew1981} are used, as well as the spacing of 0.13 \AA{} in the real space method. The simulation cell is constructed
from the spheres of the $4$ \AA{} radii around atoms.
The TDLDA approximation in the pseudopotential TDDFT method shows a very good stability and produces results in an agreement with experiments, even for large biological molecules \citep{Lopez2003}. It has been found that the results of TDDFT calculations (within the Tamm-Dancoff approximation and
using the BLYP functional) for the excitation energies of several PAH cations
agree with experiments within 0.3 eV \citep{Hirata1999}.
We have found that calculations with the B3LYP functional (still in a development in the latest version of the Octopus code) do not change substantially the spectrum of proflavine \citep{Bonaca2010}. It has also been found that spectra of proflavine obtained using the time propagation method in the Octopus code agree with results
of the Lanczos chain TDDFT module in the Quantum ESPRESSO \citep{Walker2006,Bonaca2010}.
It is known that the real-time propagation method of the Octopus code produces the whole optical spectrum up to the far-UV \citep{Castro2006}.
However, for energies above $10$ eV only envelopes of the spectra are accurate. 
In addition, there is little astronomical interest for a high-energy region of the spectrum. 
To facilitate a comparison with other studies
of PAHs and organic molecules of astrophysical interest  carried out by the same TDDFT method \citep{Malloci2007a,Malloci2007b,Bonaca2010,Puletti2010}
we show optical spectra up to {\bf 6} eV.

The hydronaphthalene cation C$_{10}$H$_9^+$ and  the dihydronaphthalene cation C$_{10}$H$_{10}^+$, with additional H atoms attached at the same positions as in Bauschlicher (1998), are studied.
Two situations of a minimal hydrogenation, where one hydrogen atom is added to the proflavine cation, are also investigated here:
an additional H atom attached as the second hydrogen atom at the central carbon atom (labelled as (I)), and at the opposite nitrogen atom in the same central ring (labelled as (II)).
We also study fully hydrogenated species of naphthalene and proflavine cations:  H$_{10}$-C$_{10}$H$_8$$^{+}$,  and H$_{14}$-C$_{10}$H$_{13}$N$_3$$^{+}$.

\section{Results and Discussion}
\label{results}

\begin{figure}
\vspace{40pt}
\centering 
\includegraphics[width=8.0cm]{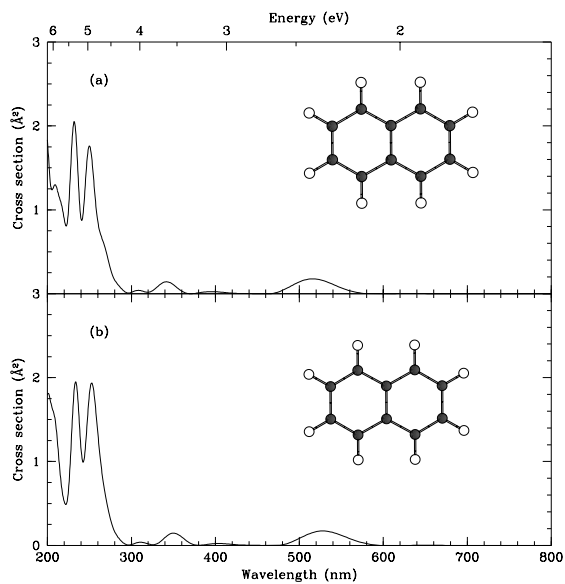}
\caption{
Optical spectra of the naphthalene cation: (a) after the geometry optimization without constraints,
(b) for the cation with the optimized structure of the neutral naphthalene.
Insets: Black balls represent carbon atoms, whereas white ones model hydrogen atoms. Structures are visualized using the XCrySDen program 
\citep{Kokalj2003}.}
\label{fig1}
\end{figure}

\begin{figure}
\vspace{50pt}
\centering 
\includegraphics[width=8.0cm]{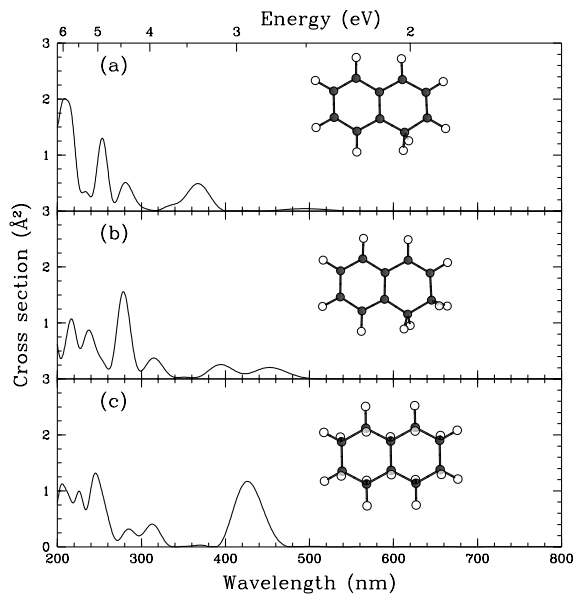}
\caption{
Optical spectra of hydrogenated species of naphthalene cation: (a) H-C$_{10}$H$_8$$^{+}$,
(b)  H$_2$-C$_{10}$H$_8$$^{+}$,
(c) H$_{10}$-C$_{10}$H$_8$$^{+}$.}
\label{fig2}
\end{figure}

\begin{figure}
\vspace{50pt}
\centering 
\includegraphics[width=8.0cm]{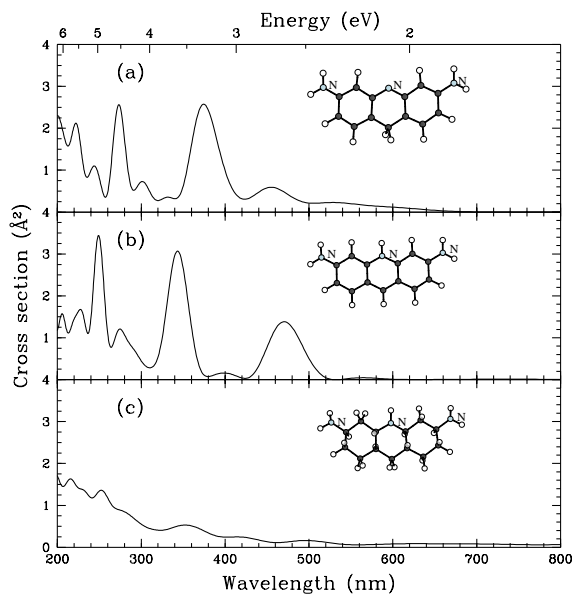}
\caption{
Optical spectra of hydrogenated species of proflavine cation: (a) H-C$_{13}$H$_{11}$N$_3$$^{+}$, additional H atom is attached
at the central C atom (position I),
(b)  H-C$_{13}$H$_{11}$N$_3$$^{+}$, additional H atom attached
at the central N atom (position II),
(c) H$_{14}$-C$_{13}$H$_{11}$N$_3$$^{+}$.
Insets: The nitrogen atoms are labelled by ``N'' on the right-hand side of corresponding balls.}
\label{fig3}
\end{figure}

The optimized structures of all investigated species of naphthalene 
and proflavine are shown as insets in Figs. 1-3.
In contrast to the naphthalene and proflavine molecules,
carbon skeletons of fully hydrogenated H$_{10}$-C$_{10}$H$_8$$^{+}$ and H$_{14}$-C$_{13}$H$_{11}$N$_3$$^{+}$ (shown in Fig.~\ref{fig2}(c) 
and Fig.~\ref{fig3}(c)) are not planar. In addition, hydrogen atoms attached in pairs on peripheral carbon atoms
are positioned below and above the middle plane of carbon rings.
Average C-C distances increase from 1.39 \AA{} in the optimized naphthalene cation to 1.51 \AA{} in  H$_{10}$-C$_{10}$H$_8$$^{+}$.
Average N-C distances increase from 1.34 \AA{} in C$_{13}$H$_{11}$N$_3$$^{+}$ to 1.41 \AA{} in H$_{14}$-C$_{13}$H$_{11}$N$_3$$^{+}$. Average C-C distances increase from 1.40 \AA{} in the proflavine cation to 1.51 \AA{} in H$_{14}$-C$_{13}$H$_{11}$N$_3$$^{+}$.
Hydrogen atoms are also attached  on the C atoms in pairs above and below rings in optimized
H-C$_{10}$H$_8$$^{+}$ (one pair, inset in Fig.~\ref{fig2}(a)) and  H$_2$-C$_{10}$H$_8$$^{+}$ (two pairs, inset in Fig.~\ref{fig2}(b)).
However, carbon skeletons are planar in both these structures with a minimal hydrogenation,
as well as in H-C$_{13}$H$_{11}$N$_3$$^{+}$ (in both positions I and II, insets in Fig.~\ref{fig3}(a) and Fig.~\ref{fig3}(b)).
The similar conclusion about a planarity of H-C$_{10}$H$_8$$^{+}$ has been reached by Alata and coworkers \citep{Alata2010a}.

Optical spectra presented in this work are calculated starting from structures optimized separately for all pure and hydrogenated cations, and without any constraints in geometrical
optimizations. We found that the fully optimized naphthalene cation is still a planar structure. However, the charge and small displacements from atom positions of the neutral naphthalene produce small differences in the spectrum. Optical spectra for the fully optimized cation and the structure where
one electron is removed from the optimized geometry of a neutral naphthalene \citep{Niederalt1995,Hirata1999,Malloci2007b}
are shown in Fig.~\ref{fig1}. The geometry used as an input for Fig.~\ref{fig1}(b) is taken from the Theoretical spectral database of PAHs \citep{Malloci2007a}.
Our result for the optical spectrum of the naphthalene cation in this situation agrees with one presented in the database \citep{Malloci2007a}.
Lines for spectra shown in Fig.~\ref{fig1}  are compared in Table~\ref{table:1}.

Optical spectra of hydrogenated protonated naphthalene species are shown in Fig.~\ref{fig2}.  
Peaks of the naphthalene cation move in  H-C$_{10}$H$_8$$^{+}$, H$_2$-C$_{10}$H$_8$$^{+}$, and H$_{10}$-C$_{10}$H$_8$$^{+}$, and their intensities change.
Optical spectra of hydrogenated protonated proflavine species are shown in Fig.~\ref{fig3}.
The most notable fact is that
the spectrum of H$_{14}$-C$_{13}$H$_{11}$N$_3$$^{+}$ (Fig.~\ref{fig3}(c)) is broad and this broadening is much more pronounced than in the fully hydrogenated naphthalene (Fig.~\ref{fig2}(c)).

Spectral lines of all naphthalene and proflavine related systems are presented in in Table~\ref{table:1} and Table~\ref{table:2}.
Visible spectra of hydrogenated naphthalene cations extend up to $\sim$ 500 nm, whereas long tails of visible spectra of hydrogenated proflavine cations
spread above 800 nm.
The photofragmentation spectroscopy measurements by Alata and coworkers
have also been shown that H-C$_{10}$H$_8$$^{+}$ absorbs in the visible, around 500 nm \citep{Alata2010a,Alata2010b}.
We also present near-UV lines (above 200 nm)
to facilitate a comparison with corresponding spectral measurements of interstellar organic materials \citep{Kwok2008}.
Such experiments are designed to be carried out on the Cosmic Origins Spectrograph (COS) of the Hubble Space Telescope \citep{Osterman2011}.
The near-UV channel of COS shows a high sensitivity in the region between 200 nm and 300 nm
where lines of hydrogenated protonated naphthalene and proflavine exist.
The lines of 281 nm and 422 nm in the spectrum of the proflavine cation [Bonaca2010] move to the 273 nm and 454 nm 
and three new lines appeared between these two
in the spectrum of H-C$_{13}$H$_{11}$N$_3$$^{+}$ (position I). 
Rather strong lines exist at 201 nm, 222 nm, and 244 nm.
An addition of the H atom to
the central N atom produces lines between 206 nm and 470 nm, as shown in Table~\ref{table:2}.
The change of optical lines is the most obvious in the spectrum of H$_{14}$-C$_{13}$H$_{11}$N$_3$$^{+}$, where the UV lines exist at 216 nm and 253 nm.
Then intensity drops, and the lines at 352 nm, 420 nm, and 496 nm are very weak.

It is known that the majority of DIBs are present in the atomic hydrogen gas \citep{Herbig1993,Snow2006}.
In Table~\ref{table:3} we compare positions of several calculated lines with the closest DIBs \citep{Hobbs2008,Hobbs2009,Tuairisg2000}.
The best agreement is for a  line of the fully hydrogenated naphthalene cation H$_{10}$C$_{10}$H$_8$$^{+}$, and a line of
H-C$_{13}$H$_{11}$N$_3$$^{+}$ (position II). Although TDDFT calculations show only trends in DIBs positions \citep{Malloci2007a,Malloci2007b,Pathak2008,Hammonds2009}, 
the agreement presented in Table~\ref{table:3} is good and deserves a further experimental investigation.
In the recent report of the high resolution spectroscopy observation of C$_{10}$H$_8$$^{+}$ in the Cernis 52,  the importance of hydronaphthalene cations  has been pointed out \citep{Iglesias-Groth2008}. Although this report has been shown to be premature
\citep{Galazutdinov2011}, the presence of the naphthalene cation and its hydrogenated forms in the interstellar medium remains an important subject
for further investigations.

\section{Conclusions}
\label{concl}

Studies of spectra of polycyclic aromatic hydrocarbon
cations with attached hydrogen atoms are important for the DIB and UIR problems, an anomalous microwave emission, as well as for general properties
of organic matter in the interstellar medium. We study the changes in electronic absorption spectra induced by
hydrogen additions to the naphthalene and proflavine cations using the time-dependent density functional method in its pseudopotential version.
Calculated spectra are based on the overall density of electronic transitions and
show that hydrogen additions substantially change intensities, shapes,
and positions of optical lines in a comparison with spectra of base cations.
Therefore, additions of hydrogen atoms is important in astrophysical applications of optical spectra of organic molecules.
Similar conclusions have been found for protonated hydrogenated coronene, ovalene, pyrene, and circumpyrene
\citep{Pathak2008,Hammonds2009}.
Calculated lines of protonated hydrogenated naphthalene and proflavine
are compared with measured DIBs. An especially good agreement exists for the visible line of the fully hydrogenated naphthalene 
cation and two visible lines of
the hydroproflavine cations.
The lines in the near-UV spectral region are also presented.
Our calculations should give guidelines for the change of near-UV and visible spectral lines for similar larger cations and their derivatives under the process of hydrogenation.

\begin{table*}
\caption{Optical spectral lines of the naphthalene cation and hydrogenated naphthalene cation species. The star labells data
for the naphthalene cation with the optimized geometry of the neutral naphthalene molecule, whereas all other data are obtained
for optimized cations. Because of a limited accuracy of TDDFT method, the calculated results are presented  rounded to whole numbers.}
\label{table:1}
\centering
\begin{tabular}{l l l l l l l}
\hline
Structure & $\lambda _1$ (nm) &  $\lambda _2$ (nm) &  $\lambda _3$ (nm) &  $\lambda _4$ (nm) & $\lambda _5$ (nm) & $\lambda _6$ (nm)\\
\hline

C$_{10}$H$_{8}$$^{+}$ &  200.  &   232. &  250. & 342.& 517. & - \\

C$_{10}$H$_{8}$$^{+}$ (*) & 201.  & 234.  & 253. & 350. & 528.& - \\

H-C$_{10}$H$_{8}$$^{+}$  & 209. & 233. & 254 & 281.& 367.& 494.\\

H$_2$-C$_{10}$H$_{8}$$^{+}$ & 217. & 238. & 279. & 315. & 395. & 453.\\

H$_{10}$-C$_{10}$H$_{8}$$^{+}$ & 206. & 226. & 246. & 285. & 313.& 426.\\
		
\hline	
\end{tabular}
\end{table*}

\begin{table*} 
\caption{Optical spectral lines of the proflavine cation and hydrogenated proflavine cation species. Labels (I) and (II) are for H atoms attached, respectively to the central carbon and central nitrogen atom. Because of a limited accuracy of TDDFT method, the calculated results are presented  rounded to whole numbers.}
\label{table:2}
{\centering
\begin{tabular}{l l l l l l l l l }
\hline
Structure & $\lambda _1$ (nm) &  $\lambda _2$ (nm) &  $\lambda _3$ (nm) &$\lambda _4$ (nm)&$\lambda _5$ (nm)&$\lambda _6$ (nm)&$\lambda _7$ (nm)&$\lambda _8$ (nm)\\
\hline

C$_{13}$H$_{11}$N$_3$$^{+}$ & 281.$^{1}$ & 422.$^{1}$ & -  &&&&&\\

H-C$_{13}$H$_{11}$N$_3$$^{+}$ (I) & 201. & 222. & 244. &273.& 301.& 332.& 375.& 454.\\

H-C$_{13}$H$_{11}$N$_3$$^{+}$ (II)& 206. & 228. & 249. &274.& 343.& 399.& 470.& -\\

H$_{14}$-C$_{13}$H$_{11}$N$_3$$^{+}$ & 216. & 253. & 352.& 420.& 496. & - & - & - \\
	
\hline	

\end{tabular}
}

$^{1}${\citep{Bonaca2010}}

\end{table*}

\begin{table*}
\caption{Lines calculated using the time-dependent density functional theory (TDDFT) method 
 compared to the nearest DIBs. Because of a limited accuracy of TDDFT method, the calculated results are presented  rounded to whole numbers.
 FWHM of experimental lines are also shown.}
\label{table:3}
{\centering
\begin{tabular}{l l l l}
\hline
Structure & $\lambda$(TDDFT) (nm)  &  $\lambda$(DIB) (nm) & FWHM (nm)\\
\hline

C$_{13}$H$_{11}$N$_3$$^{+}$ & 422.$^{1}$ & 417.55$^{2}$ &  1.72\\

H-C$_{13}$H$_{11}$N$_3$$^{+}$ (I) & 454. & 450.17$^{3}$ & 0.30\\

H-C$_{13}$H$_{11}$N$_3$$^{+}$ (II) & 470. & 476.26$^{3}$ & 0.25\\

H$_{10}$C$_{10}$H$_8$$^{+}$ & 426.  & 425.90$^{4}$ &  0.11\\

\hline	
\end{tabular}
}

$^{1}${\citep{Bonaca2010}}

$^{2}${\citep{Tuairisg2000}}

$^{3}${\citep{Hobbs2009}}

$^{4}${\citep{Hobbs2008}}
\end{table*}

\section*{Acknowledgments}
This work has been done under the HR-MZOS project 119-1191458-1011, and using computer resources at
the University of Zagreb Computing Centre SRCE. The authors would like to thank
Alberto Castro and Layla Martin-Samos for discussions. 

\bibliographystyle{mn2e} 
\bibliography{hpfnaf}

\label{lastpage}
\end{document}